# The feasibility of combining communication BCIs with FES for individuals with locked-in syndrome

Evan Canny[1], Mariska J. Vansteensel[1], Sandra M.A. van der Salm[1], Gernot R. Müller-Putz[2], Julia Berezutskaya[1]*

[1]Brain Center, department of Neurology and Neurosurgery, University Medical Center Utrecht, Utrecht, the Netherlands

[2]Institute of Neural Engineering, Laboratory of Brain-Computer Interfaces, Graz University of Technology, Graz, Austria

Corresponding author: y.berezutskaya@umcutrecht.nl

**Abstract**

Individuals with a locked-in state live with severe whole-body paralysis that limits their ability to communicate with family and loved ones. Recent advances in the brain-computer interface (BCI) technology have presented a potential alternative for these people to communicate by detecting neural activity associated with attempted hand or speech movements and translating the decoded intended movements to a control signal for a computer. A technique that could potentially enrich the communication capacity of BCIs is functional electrical stimulation (FES) of paralyzed limbs and face to restore body and facial movements of paralyzed individuals, allowing to add body language and facial expression to communication BCI utterances. Here, we review the current state of the art of existing BCI and FES work in people with paralysis of body and face and propose that a combined BCI-FES approach, which has already proved successful in several applications in stroke and spinal cord injury, can provide a novel promising mode of communication for locked-in individuals.

# Introduction

Locked-in individuals have a neurological impairment that leads to severe whole-body paralysis with preserved consciousness and cognitive functioning (American Congress Of Rehabilitation Medicine, 1995). *Locked-in syndrome* (LIS) can be caused by various conditions, including brainstem stroke, trauma or a progressive neurodegenerative disease (e.g., amyotrophic lateral sclerosis; ALS). Depending on the severity of the paralysis, researchers identify three types of LIS (Bauer et al., 1979). *Classical LIS* is characterized by paralysis of all four limbs (quadriplegia), bilateral facial paralysis, and loss of voice and speech (aphonia), while retaining the ability to produce eye movements (E. Smith & Delargy, 2005). *Incomplete LIS* is characterized by remnants of voluntary movements other than eye movements. *Total LIS* is characterized by whole body paralysis including the eye muscles, causing total immobility and inability to communicate. While individuals with brainstem stroke or trauma either retain a static LIS state or even recover from it (Patterson & Grabois, 1986), individuals with a neurodegenerative disease may gradually progress through different LIS stages over time, in some cases ultimately transitioning towards a total whole-body paralysis with no means to communicate with their family, friends and caregivers (Nakayama et al., 2016).

A crucial factor in determining the quality of life of locked-in individuals is their ability to communicate with family and loved ones (Bruno et al., 2011; Rousseau et al., 2015). Bruno et al. (2011) showed the inability to speak in locked-in individuals with LIS due to brainstem stroke is associated with poor subjective wellbeing. Individuals with classical LIS can typically use assistive communication technology based on residual muscle movement, for example, vertical eye movements and eyeblinks (Bruno et al., 2011). Individuals with total or near-total LIS are not able to use residual movements reliably, and inability to use assistive technology for communication significantly lowers the quality of life (Rousseau et al., 2015).

An alternative to assistive communication devices based on residual muscle movement is *brain-computer interface* (BCI) technology. Both non-implantable and implantable BCI systems record brain signal changes associated with cognitive processing or attempts to produce functional movements (e.g., attempt to move a limb, to make a facial expression or to speak) and use them to control a computer, robotic arm, or another external device. Identifying and directly connecting neural signatures of intended movements to external output systems opens up a possibility for the locked-in individuals to communicate while bypassing their impaired motor pathways (Birbaumer, 2006; Brumberg et al., 2010; Lebedev & Nicolelis, 2006; Leuthardt et al., 2009; Wolpaw et al., 2002; Wolpaw & Wolpaw, 2012). Several studies have demonstrated the potential of BCI to provide communication in persons with LIS (Birbaumer et al., 1999; Chaudhary et al., 2022; Holz et al., 2015; Kübler et al., 2009; Milekovic et al., 2018; Oxley et al., 2021; Vansteensel et al., 2016).

Using decoded movement attempts of locked-in individuals to communicate via external output systems is not the only promising BCI avenue. Studies on individuals with stroke and spinal cord injury resulting in severe body paralysis have shown that BCI combined with *functional electrical stimulation* (FES) can create an electronic neural bypass of damaged neural pathways by using neural signals to drive stimulation of peripheral nerves and thereby trigger paralyzed muscle activation (Ajiboye et al., 2017; Bouton et al., 2016; Muller-Putz et al., 2019; Pfurtscheller et al., 2003).

Given the fast progress being made in the fields of communication BCIs and BCI-FES, here, we argue that it is worthwhile to investigate whether these concepts can be combined in order to enrich the communication abilities of persons with LIS. In this previously unexplored avenue of research, the application of BCI-triggered FES could be extended to facial muscles to restore simple movements, such as eye blinks or finger taps, or more complex movements, such as facial expressions and potentially even vocalizations and speech articulation.

In the present review, we assess feasibility of the FES technology to boost existing communication BCI approaches for locked-in individuals. We review existing work on assistive communication technology for locked-in individuals, including non-implantable and implantable BCIs, discuss studies on FES applied to body and face muscles and overview existing developments on combining BCI and FES. Finally, we discuss the potential, challenges, and possible future directions of implementing BCI-FES systems for the restoration of communication function in locked-in individuals.

# BCI communication for locked-in individuals

## Augmentative and alternative communication devices

Studies on augmentative and alternative communication devices typically distinguish three types of such technology: no-tech, low-lech and high-tech devices (Fried-Oken et al., 2015). No-tech devices rely on body movements and residual speech without aid of additional devices such as computers or eye trackers. Individuals with classical and incomplete LIS, who retain the ability to produce eye movements, can use no-tech "yes-no" response strategy with eyeblinks to answer closed questions or spell with the help of a communication partner reciting letters out loud. Other residual movements such as chin, jaw, forehead, head movements, remaining vocalizations and their combinations can also be employed (Bauby, 2008; Feldman, 1971; Gauger, 1980). Low-tech devices rely on non-computer based equipment such as pen and paper or communication boards with letters, words or pictures. Many systems employ low-tech solutions that combine "yes-no" strategy and communication boards with eye movements and eyeblinks (Gallo & Fontanarosa, 1989; Kenny & Luke, 1989). High-tech devices use computer-assisted technology including audio synthesis of selected and typed text. Depending on the strength and reliability of residual movements, these systems can either facilitate the use of alphabet spelling and pictogram communication boards with automatic selection scrolling (switch scanning) or provide more elaborate text-to-speech solutions based on mouse or joystick control with residual movements. Other high-tech systems can rely on gaze control of a computer cursor based on eye-tracking recordings (L. J. Ball et al., 2010; Linse et al., 2018).

The described systems are used extensively with locked-in individuals who retain a certain level of reliable voluntary muscle control (Lugo et al., 2015; Snoeys et al., 2013). Unfortunately, these systems gradually become unreliable in LIS caused by ALS and can be unreliable after total LIS onset caused by other conditions (Hinterberger et al., 2005; Murguialday et al., 2011). Inability to use assistive communication technology based on residual movement may be related to difficulty maintaining stable head position (Spataro et al., 2014), progressive oculomotor impairment and eye gaze fatigue (Averbuch-Heller et al., 1998; Nakayama et al., 2016; Shaunak et al., 1995), pupil dilation due to the use of medication (S.-H. K. Chen & O'Leary, 2018) and other reasons. In such cases, BCIs can be a promising alternative means of communication for the affected individuals (Vansteensel et al., 2023).

## Non-implantable BCI for communication in LIS

Non-implantable neural recording modalities include scalp electroencephalography (EEG), magnetoencephalography (MEG), functional near-infrared spectroscopy (fNIRS) and functional magnetic resonance imagining (fMRI). Practically, scalp EEG and, to some extent, fNIRS are more widely used in BCI research given their portability and affordability compared to MEG and fMRI. Scalp EEG setups can vary in design and number of electrodes depending on the clinical, research or commercial use. For BCI applications, EEG caps of 16 to 128 electrodes covering the entire brain are typically used. Scalp EEG records electrical neural signals at a high temporal resolution but suffers from low spatial resolution and specificity, where the recorded signals typically represent a mixture of neural activity from widespread brain areas resulting from passing of the signals through the protective layers of the brain, skull, and scalp. The working principle of fNIRS is similar to fMRI – the technique measures local changes in blood flow around the brain surface associated with changes in cognitive activity. Because blood flow changes are slow, fNIRS suffers from low temporal resolution. Signal quality of both EEG and fNIRS is affected by motion-related and other artifacts. Moreover, continuous wearing of the EEG or fNIRS cap is not practical for 24/7 home use (Zickler et al.,

2011), and in most current setups requires subject and session specific recalibration (Lotte et al., 2018).

A number of strategies have been employed for decoding EEG brain activity for BCI communication, including decoding of mental effort, motor or speech imagery and movement attempts. Use of mental effort is based on a premise that reliable decoding of any meaningful cognitive effort, not specifically related to motor activity or speech, can be used for communication. In that regard, an important subset of early EEG-based BCI research focused on evoked potentials – time-domain EEG signatures of task-related activity. This includes steady-state visually evoked potentials (SSVEP) – a visual cortex response to frequency and position of flickering visual stimuli (Adrian & Matthews, 1934; Regan, 1989); P300 – cortical responses to infrequent stimuli, and motor evoked potentials (Farwell & Donchin, 1988). P300 has been widely used for spelling and choice-based communication applications and has been adapted to various modalities: visual, auditory and vibrotactile (Kaufmann et al., 2014; Kübler et al., 2009; Nijboer et al., 2008). Frequency-domain EEG signatures widely used in BCI, on the other hand, include slow cortical potentials – slow (below 1 Hz) widespread positive and negative signal amplitude shifts related to task activity (Birbaumer et al., 1990) and sensorimotor rhythms in mu and beta frequency bands modulated by executed, imagined or attempted movement (Obermaier et al., 2003; Yuan & He, 2014). Another venue of research is based on error potential responses associated with processing of errors (Falkenstein et al., 1995). Error potentials can complement other paradigms, such as P300-based BCIs, wherein P300 responses indicate decoding targets and error potentials arise as response to decoded misses.

Decoding motor and speech imagery with EEG was based on the idea that intention to communicate in BCI target users results in neural processes similar to those underlying imagery in able-bodied individuals. Previous work has shown successful decoding of imagined movements and speech information from EEG (Deng et al., 2010; Min et al., 2016; Neuper et al., 2009; Pfurtscheller et al., 2006; L. Wang et al., 2013, see Lopez-Bernal et al., 2022 for reviews). In the recent years, however, the BCI field has been focusing more on decoding of actual movements and speech recognizing that attempted movements of BCI users may have more in common with actual movements of able-bodied subjects than motor imagery (Hotz-Boendermaker et al., 2008), which in its turn may involve specific and distinct neural mechanisms. Several studies have demonstrated decoding produced speech and movement information from EEG (Bradberry et al., 2010; Liao et al., 2014; Ofner & Muller-Putz, 2012; Sharon et al., 2020; Waldert et al., 2008). Multiple reviews cover EEG applications for BCI (Abiri et al., 2019; B. He et al., 2015; Lazarou et al., 2018; Norcia et al., 2015; Rashid et al., 2020; Värbu et al., 2022; Yuan & He, 2014).

Work on fNIRS has so far been more limited but already shown successful decoding of motor, mental workload (Herff et al., 2014), speech-related (Cao et al., 2021; Herff et al., 2012; Liu & Ayaz, 2018) and other task-related (Nagels-Coune et al., 2020) information from neural signals.

Many of the proposed paradigms have been tested in individuals with LIS, including SSVEP (Combaz et al., 2013; Lesenfants et al., 2014), P300 (Guger et al., 2017; Pokorny et al., 2013), slow cortical potentials (Kübler et al., 2001), sensorimotor rhythms (Kübler et al., 2005) and error potentials (Spüler et al., 2012). Limited testing with locked-in individuals was performed with fNIRS (Abdalmalak et al., 2017; Gallegos-Ayala et al., 2014). Overall, the performance of non-implantable BCIs is promising for a number of tasks but varies greatly across individuals. Some research, however, indicates that prolonged use of paradigms based on SSVEPs and P300 may be tiresome and time-consuming (Pires et al., 2012). Compared to naturalistic verbal and non-verbal communication, these paradigms may not be entirely intuitive and comfortable for users. Importantly, gaze-based BCI control paradigms, such as P300, may not work long-term for LIS caused by ALS (Brunner et al., 2010; Hinterberger et al., 2005; Murguialday et al., 2011; Vansteensel et al., 2023), as oculomotor activity is progressively affected by the disease. Instead, BCI paradigms based on attempted movement and speech that do not rely on residual muscle activity, may be more natural and intuitive, and therefore preferred strategies for BCI-based communication (Branco et al., 2021).

**Implantable BCI for communication in LIS**

Implantable BCIs for LIS individuals capitalize on superior signal quality provided by direct recordings from the brain tissue and the high temporal and spatial resolution that implanted neural recording techniques provide compared to EEG, MEG and fNIRS. Specifically, intracranial EEG that includes electrocorticography (ECoG), or surface grid electrodes, and stereo-EEG (sEEG), or depth electrodes, allows recording from 1 to 2.3 mm diameter patches of cortex around each electrode and can sample from up to 256 electrodes. ECoG provides a spatial resolution of 3-10 mm with gaps between electrodes and typically requires an invasive surgical procedure, yet signals remain stable over time (Larzabal et al., 2021), and there is no tissue scarring. Therefore, ECoG has been tested and approved for long-term use in human participants of BCI trials. Related stentrode technology records brain signals via electrode arrays arranged on a flexible mesh tube (Oxley et al., 2016, **Figure 1**). An array of up to 16 stentrodes, 750 μm diameter each, can be inserted through a jugular vein and guided towards a vessel adjacent to sensorimotor brain area with a catheter. Multielectrode arrays (MEAs) and neurotrophic electrodes allow individual neural cell recordings. While providing superior spatial resolution, MEAs offer smaller cortical coverage with up to 256 needles per array covering up to 2.8 x 2.8 mm of cortex. MEAs are made of stiff materials and penetrate the cortex, which can lead to tissue scarring and gradual signal loss (Colachis et al., 2021). Typically, high frequency (> 60 Hz) and low frequency (13-30 Hz) components of ECoG and stentrode signals are used for decoding information from the brain. In MEA, spike information and spike band power are used.

Various neural decoding strategies have been explored in pre-clinical studies using invasive recordings in non-human primates and able-bodied individuals. Non-human primate studies with ECoG and MEAs have targeted decoding of reaching and grasping hand movements as well as continuous hand movement trajectories for potential translation to human BCIs (Carmena et al., 2003; Serruya et al., 2002; D. M. Taylor et al., 2002; Velliste et al., 2008). Human recordings with temporarily implanted ECoG or sEEG electrodes are possible in patients who undergo clinical epilepsy monitoring. In such cases, electrode implantation can help localize epileptogenic sources and guide subsequent resection of the affected neural tissue. Research in this setting has contributed to detailed study of sensorimotor brain areas, such as the "hand knob" and "face area", revealing somatotopic organization of individual fingers (Schellekens et al., 2018) and speech articulators (Bouchard et al., 2013), respectively.

In addition, this pre-clinical human research has demonstrated successful decoding of discrete motor information, such as basic hand movements (Pistohl et al., 2012; Spüler et al., 2014), hand movement towards a specific location (Bundy et al., 2016; W. Wang et al., 2013), hand gestures (Bleichner et al., 2016; Branco et al., 2017; Li et al., 2017; Pan et al., 2018), basic mouth movements (Salari et al., 2019), facial expressions (Salari et al., 2020) and continuous motor information, including 2D and 3D finger trajectories, speed, acceleration and force of movement (Bouton et al., 2016; Bundy et al., 2016; Collinger et al., 2012; Śliwowski et al., 2022; Willett et al., 2021). In parallel, decoding of speech has shown success in decoding individual words (Berezutskaya, Freudenburg, et al., 2022; Kellis et al., 2010; Martin et al., 2016; Metzger et al., 2022; Moses et al., 2021; Stavisky et al., 2019), phonemes (Blakely et al., 2008; Bouchard & Chang, 2014; Herff et al., 2015; Mugler et al., 2014; Ramsey et al., 2018), syllables (Livezey et al., 2019; Stavisky et al., 2019) and even continuous decoding of full phrases and sentences (Anumanchipalli et al., 2019; Herff et al., 2015, 2019; Makin et al., 2020; Moses et al., 2021; Sun et al., 2020; Willett et al., 2023).

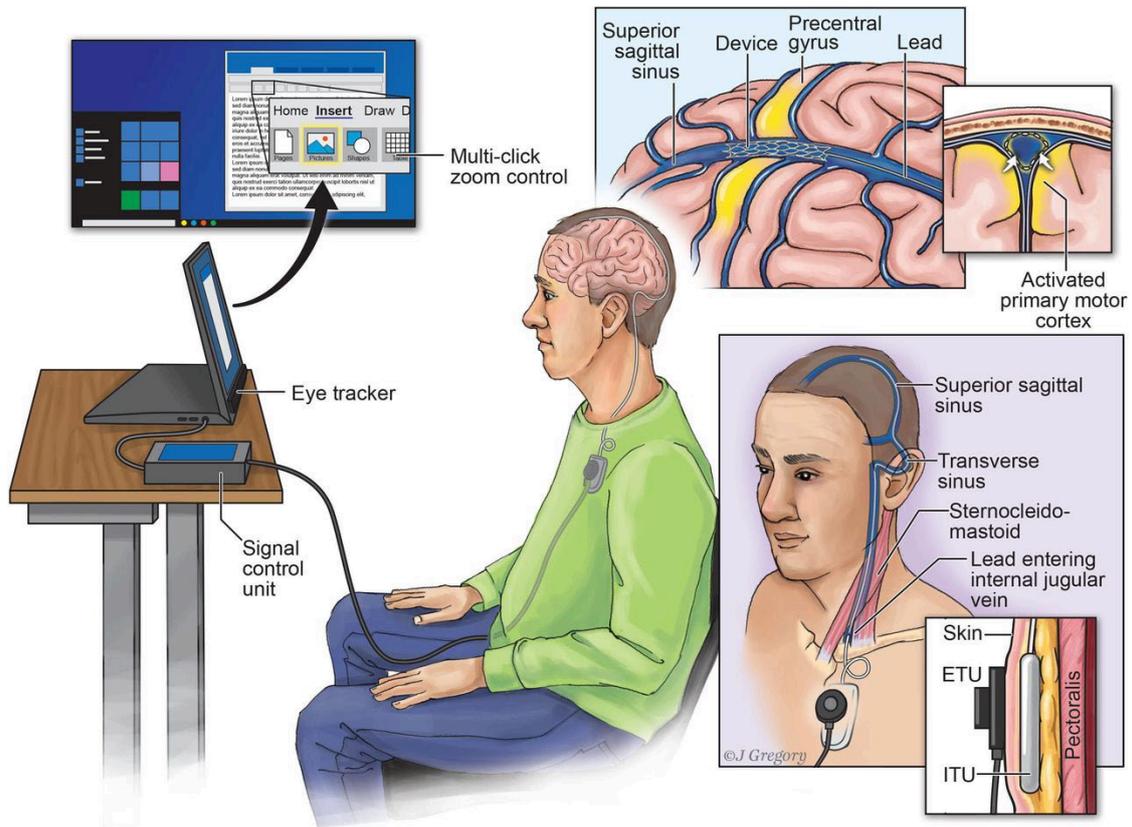

**Figure 1**. Example of an implantable BCI system for a locked-in individual with ALS based on attempted hand movements. Copyright: (Oxley et al., 2021)

Success of pre-clinical studies with non-human primates and able-bodied humans led the way towards a demonstration of a proof-of-concept BCI system that could enable communication for locked-in individuals (Vansteensel et al., 2016). The study demonstrated an ECoG-based BCI for computer control and spelling based on accurate decoding of a finger tapping movement attempted by a user with ALS. Later studies have shown similar success using stentrodes and decoding of attempted leg movement (Mitchell et al., 2023; Oxley et al., 2021). MEA-BCI systems have been employed for decoding of hand movement "point and click" patterns for computer cursor control by participants with severe paralysis (Bacher et al., 2015; Pandarinath et al., 2017). Another recent study trained a user with total LIS to spell letters using a neurofeedback strategy: the user's neural activity levels were mapped to auditory tones and played back to the user, who was asked to modulate his brain activity to match the target tone (Chaudhary et al., 2022). The most recent BCI work with both ECoG and MEAs in people with motor impairment demonstrated successful attempted word and speech decoding for display on a computer screen for communication (Metzger et al., 2022; Moses et al., 2021; Willett et al., 2023). Importantly, only a handful of studies investigated the potential of implanted BCIs in individuals with late-stage ALS in total or near-total LIS, including complete inability to vocalize (Chaudhary et al., 2022; Vansteensel et al., 2016). For translation of the BCI technology from research to real-world user applications, fully-implantable home-use BCI systems will be needed, wherein BCI users can control the computer with their brain activity continuously throughout day and night without supervision by researchers (Oxley et al., 2021; Vansteensel et al., 2016).

Overall, implantable BCI approaches offer several attractive features and have shown high performance when used in naturalistic communication paradigms based on attempted movements and speech. At the same time, non-implantable BCI systems can also reliably decode cognitive and motor attempts that can be used for communication while offering inexpensive portable BCI solutions without complex surgical intervention. In both implantable and non-implantable BCI setups, decoded communication attempts can be displayed on a computer, used for brain-driven control of a cursor, a custom or commercial speller or communication app. External devices and tools can also be used to augment the user experience. These can involve a speech synthesis device, a virtual avatar, an external mobility device, such as a robotic arm or exoskeleton (Benabid et al., 2019; Downey et al., 2016; Hochberg et al., 2006; Wodlinger et al., 2015; see Bockbrader, 2019 for a review). Among such tools used in BCI applications for rehabilitation in stroke and spinal cord injury is also functional electrical stimulation (FES). Similar to the use of external mobility devices, FES is aimed to restore or improve BCI user's motor control. It does so by reanimating weakened or paralyzed muscles of user's body and limb. With the majority of FES applications being directed to people with stroke or spinal cord injury, it has not yet been employed directly for communication purposes.

# Functional electrical stimulation (FES)

## What is FES?

FES refers to application of electrical stimulation to neural tissue with the aim to restore its lost or damaged function (Peckham & Knutson, 2005; Rupp, 2021; Rushton, 1997). Often, FES is applied to peripheral nerves to generate contraction of the connected muscle and induce functional movement. The electrical stimulation can be administered using non-invasive setups (*transcutaneously*, with electrodes and stimulator on the surface of the skin) or invasive setups (*percutaneously*, with electrodes piercing the skin and an external stimulator, or via *fully-implantable* systems with electrodes and a stimulator implanted under the skin).

Transcutaneous FES is carried out with electrodes placed on the surface of the skin. It is a non-invasive, relatively inexpensive, low maintenance and widely popular approach in clinical and research settings. However, because there are different structures between the surface electrode and the stimulation target, including pain fibers, other non-target nerves and muscles, it may be difficult to use transcutaneous FES for activation of isolated muscles, especially deep muscles. As a result, stimulation typically requires higher charges compared to invasive techniques. Since pain fibers can be in the way of stimulating the target muscle, unlike invasive approaches, surface stimulation can cause skin sensations that can range from slightly tingling to strong discomfort and painful (Ilves et al., 2020; Peckham & Knutson, 2005).

FES with percutaneous electrodes involves applying stimulation directly to the nerve in the muscle, which can be achieved using needle electrodes that pierce the skin (Marquez-Chin & Popovic, 2020; Masani & Popovic, 2011; Peckham & Knutson, 2005; Quandt & Hummel, 2014; Ragnarsson, 2008). Percutaneous electrodes are less likely to cause pain as they bypass the skin and can activate isolated and deep muscles. In FES setups with percutaneous electrodes, the stimulation device is placed externally. Fully-implantable FES systems, on the other hand, were designed for long-term use and involve under-the-skin implantation of the electrodes as well as the stimulator device that can be powered and controlled with an external unit. Cochlear implants that restore hearing and implants for restoring grasp in tetraplegia are examples of implantable FES systems.

Typically, FES is applied to the point where the nerve enters the target muscle – the motor point, as stimulating the motor point requires the lowest stimulation threshold to induce muscle contraction (Rupp, 2021). Implantable FES systems may allow for more fine-grained stimulation targeting and can include electrode placement on the muscle surface, within a muscle, adjacent to a nerve, or around a nerve (Peckham & Knutson, 2005). In most cases, these different configurations aim at exciting the nerves that innervate the muscles. Some literature refers to direct stimulation of muscles in the case of severe muscle denervation (see **Challenges and future directions for more detail**).

Apart from the differences in the outlined FES setups, other factors can influence FES outcomes and performance including conductivity of the underlying tissue, muscle training and the degree and treatment of muscle fatigue (Peckham & Knutson, 2005; Rupp, 2021). Moreover, FES is characterized by a number of stimulation parameters (**Figure 2**), such as the stimulus waveform, the pulse duration, or width, the current amplitude, and the stimulation frequency, which can all affect stimulation results. Biphasic waveforms are deemed safer for the underlying nervous tissue as they can balance out the electrochemical processes caused by stimulation thereby minimizing tissue damage (Mortimer, 1981). Some research has examined the effects of distinct waveform shapes and other parameters on FES performance and user comfort (Baker et al., 1988; Bennie et al., 2002; Bowman & Baker, 1985; Crago et al., 1974;

Ilves et al., 2020; Mäkelä et al., 2020), but overall such effects are non-trivial to compare across studies due to the large degree of variability in FES applications.

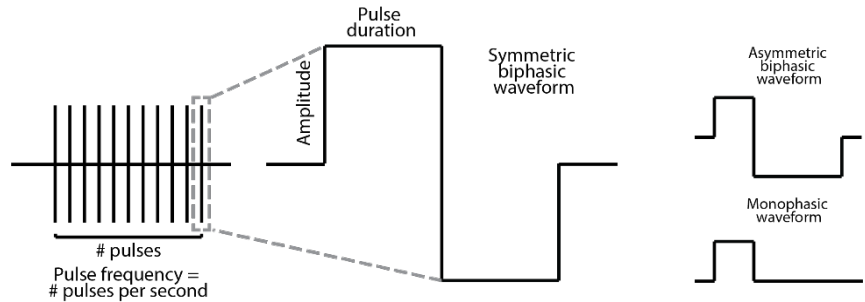

**Figure 2.** A visual illustration of the stimulation parameters used in FES research including pulse frequency, pulse amplitude, pulse duration and different types of waveform: symmetric biphasic, asymmetric biphasic and monophasic waveform.

As such, FES applications vary with respect to the area of stimulation (face, hand, leg, ear, bladder, etc.), specific underlying neural impairment (stroke, spinal cord injury, deafness, etc.), target application (restoration of respiration, grasp movements, walking, hearing, etc.), design of the control system (user-controlled open-loop systems, systems with continuous stimulation, closed-loop or feedforward control, etc.). Multiple reviews cover FES principles and applications (Eck & Rupp, 2021; Marquez-Chin & Popovic, 2020; Masani & Popovic, 2011; Peckham & Knutson, 2005; Popović, 2014; M. R. Popovic et al., 2016; Prodanov et al., 2003; Rupp et al., 2015; Sheffler & Chae, 2007; Zhang et al., 2007), including for stroke (Alashram et al., 2022; Howlett et al., 2015; Kottink et al., 2004; Quandt & Hummel, 2014; Vafadar et al., 2015), spinal cord injury (Braz et al., 2009; Gater et al., 2011; Ho et al., 2014; Luo et al., 2020; M. R. Popovic et al., 2001; Ragnarsson, 2008), Parkinson's disease (Sujith, 2008) and facial nerve paralysis (Burelo-Peregrino et al., 2020; Fargher & Coulson, 2017).

**FES in limb, body, and face paralysis**

Given the focus of the present paper on communication for individuals with severe whole-body paralysis, below we review studies on clinical and research FES applications that can induce higher-level voluntary motor activity, such as functional movements of body, limbs, and face. Many people with LIS use residual movements of the hand or face to answer closed questions or to control their communication technology device. We propose that several types of FES-generated movements can potentially be used either for coded communication, similar to augmentative and alternative communication devices based on residual movements, for example, eye blinks, finger taps or other upper or lower limb movements, or for direct verbal or non-verbal communication, for example, facial expressions, sign language or speech.

*Limb and body paralysis*

The utilization of FES is well-established within research and clinical settings for the improvements or restoration of motor functioning following several medical conditions, for example, stroke, spinal cord injury, cerebral palsy, multiple sclerosis, Parkinson's disease, trauma, and others. Many of these conditions result in weakness or paralysis of different body parts, which can affect functional upper and lower limb movements and balance.

Most work on inducing or strengthening upper limb movements with FES has been done with transcutaneous electrodes (except for studies that use the implanted Freehand system). Previous work has shown that FES delivery to peripheral nerves can facilitate and induce various functional movements of hand and arm (Peckham & Knutson, 2005; Rupp et al., 2015). Concurrent transcutaneous stimulation in large shoulder and arm muscles, such as anterior and posterior deltoids, triceps and biceps, can induce reaching movements in able-bodied participants (Resquín et al., 2016; Westerveld et al., 2014) and individuals with upper limb paralysis (Thrasher et al., 2008).

Different types of hand grasp, including the palmar power grasp used to hold large, heavy objects by flexing fingers against the palm, and the pinch precision grasp used to hold objects between the thumb and other fingers for fine-grained manipulation (Feix et al., 2016), can be improved and induced by transcutaneous FES in able-bodied participants (Vinil et al., 2014) and individuals with upper limb paralysis (Thrasher et al., 2008). This can be done by concurrent stimulation in forearm muscles, such as wrist and finger flexion muscles (flexor carpi radialis, flexor carpi ulnaris, flexor digitorum profundus, flexor digitorum superficialis and others), and wrist and finger extension muscles (extensor carpi radialis, extensor carpi ulnaris, extensor digitorum and others). Implanted FES has been shown to induce grasp movements in individuals with paralysis (Peckham et al., 2001).

In addition to basic grasp movements, transcutaneous FES has been shown to induce finger extension in the paralyzed arm (Cauraugh et al., 2000; Y. Chen et al., 2019; Knutson et al., 2012); hand opening, fist clenching and wrist extension with the intact and paretic arm (Y. Chen et al., 2019; S. He et al., 2023); functional gestures such as closing a drawer, button pressing, switching on a light switch in both able-bodied participants and individuals with upper limb paralysis (Kutlu et al., 2016) and individual finger movements and hand postures for playing a musical instrument in an able-bodied participant (Tamaki et al., 2011).

Transcutaneous and implantable FES can also facilitate and induce movements of the lower limb including movements of the thigh, ankle and foot by stimulating in hamstring, quadricep, peroneus longus and other leg muscles (Bajd et al., 1983; Kern et al., 1999; Kuzelicki et al., 2002; Thrasher et al., 2006).

Stroke and spinal cord injury have a long history of using FES-based therapy in a clinical setting with commercially developed FES systems with transcutaneous stimulation, such as the NESS Handmaster (Alon & Ring, 2003; Ring & Rosenthal, 2005), the H200 Wireless Hand Rehabilitation System (Alon et al., 2007), the COMPEX Motion Stimulator (M. R. Popovic et al., 2006; Thrasher et al., 2008), MyndMove (Hebert et al., 2017), Parastep-1 (Graupe & Kohn, 1994) and implanted setups, such as the Freehand System (B. Smith et al., 1998). FES-based therapy been shown to significantly improve upper and lower limb motor function in both stroke (Dunning et al., 2015; Hebert et al., 2017; Marquez-Chin & Popovic, 2020; M. B. Popovic et al., 2004; Stein et al., 2010; P. N. Taylor et al., 1999; Thrasher et al., 2008) and spinal cord injury (Johnston et al., 2003; Kapadia et al., 2013; Keith et al., 1989; M. Kim et al., 2004; Mangold et al., 2005; Peckham et al., 1980; M. R. Popovic et al., 2011; Rupp et al., 2015; P. Taylor et al., 2002; Thrasher et al., 2006). The knowledge and experience about FES, its clinical practice and commercial applications accrued in the field of motor restoration after stroke and spinal cord injury can be particularly informative for the development of stimulation-based applications in novel fields, such as FES for communication in individuals with LIS.

*Facial nerve paralysis*

Another type of FES application for inducing functional movements that could be used for communication is work on facial nerve paralysis. Facial paralysis can occur as a result of damage to the facial nerve or facial muscles caused by various conditions. Unilateral facial nerve paralysis is rather common and is typically caused by Bell's palsy, infection, trauma, stroke, developmental condition or tumor (Owusu et al., 2018). Bilateral facial nerve paralysis is rarer and can be caused by the same conditions as well as the Moebius syndrome. Facial nerve paralysis affects 20 to 30 per 100,000 individuals each year across several countries (Alanazi et al., 2022; Cha et al., 2008; Junior et al., 2009; Peitersen, 2002).

The facial muscles (**Figure 3**) are innervated by the facial nerve (VII cranial nerve) and are responsible for essential functional movements of the face. Transcutaneous FES administration in the eye muscle (orbicularis oculi) has been shown to produce eyeblinks at low levels of pain and discomfort in able-bodied participants (Frigerio & Cavallari, 2012; Ilves et al., 2019; Lylykangas et al., 2018; Rantanen et al., 2016) and in individuals with facial nerve paralysis (Frigerio et al., 2015; Lylykangas et al., 2020; Mäkelä et al., 2021; Raslan et al., 2020).

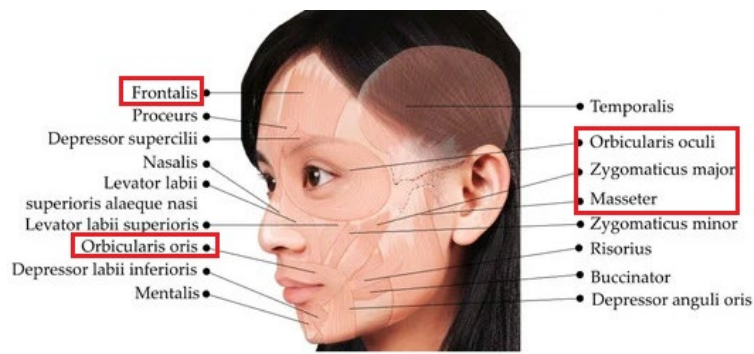

**Figure 3.** Illustration of the facial muscles. The facial muscles that have been stimulated in previous facial FES studies are highlighted in red: the frontalis (forehead), orbicularis oculi (eye), zygomaticus major (cheek), masseter (cheek), and orbicularis oris (mouth). Adapted from (Wu et al., 2022), copyright by the authors.

Recent work has demonstrated that transcutaneous FES can induce eyebrow raises in able-bodied participants (Ilves et al., 2019) and in individuals with facial nerve paralysis (Mäkelä et al., 2019, 2020) by triggering forehead muscle (frontalis) contraction (**Figure 4**). Transcutaneous FES in the mouth muscle (orbicularis oris) can induce lip puckering in able-bodied participants (Ilves et al., 2019) as well as in individuals with facial nerve paralysis (Mäkelä et al., 2019; Raslan et al., 2020; Volk et al., 2020). Cheek muscle (zygomaticus major) activation with transcutaneous FES can induce a smile in able-bodied participants (Ilves et al., 2019) and in individuals with facial nerve paralysis (Arnold et al., 2021; Mäkelä et al., 2019; Raslan et al., 2020; Volk et al., 2020). Inducing a smile with transcutaneous FES has however been more challenging compared to eyeblinks, eyebrow raises and lip pucker due to cross-stimulation of the eye and increased levels of discomfort (Ilves et al., 2019). In all other cases, studies reported low levels of user pain and discomfort in both able-bodied participants and individuals with facial paralysis.

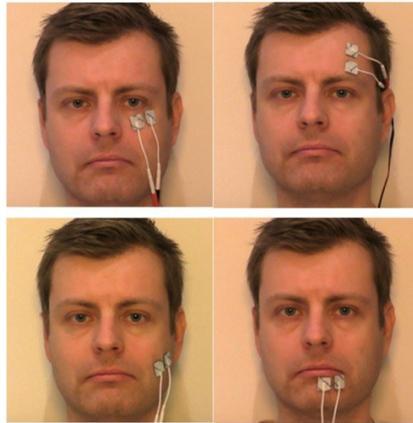

**Figure 4.** Example of applying FES electrodes for external activation of facial muscles. Copyright: (Ilves et al., 2019).

Percutaneous stimulation in facial muscles of able-bodied individuals has been shown to trigger appearance of fine-grained facial movements, such as inner and outer eyebrow raises, cheek raising, eyebrow lowering, smiling, lip stretching, chin raising, nose wrinkling and others (Waller et al., 2006). Percutaneous FES in the temporarily anesthetized forehead muscle of an able-bodied participant induced bilateral eyebrow raises (Kurita et al., 2010). In individuals with facial nerve paralysis, intraoperative transcutaneous and percutaneous stimulation in the chewing muscle (temporalis) has been shown to produce a smile (Har-Shai et al., 2010). Overall, however, despite these promising results, research on facial stimulation with percutaneous and implanted electrodes appears to be rather scarce.

Finally, a recent study showed that some speech-related activity can be induced in an able-bodied participant by application of transcutaneous FES in the cheek muscle (zygomaticus major, Schultz et al., 2019). The study demonstrated that FES could induce pronunciation of a consonant /v/ with acoustic sound features similar to those of the consonant /v/ freely produced by the participant without applying stimulation. This result, however, could only be achieved with self-controlled stimulation by the participant but not with externally controlled stimulation by the experimenter.

FES-based therapy is an actively developing field that aims to alleviate physical consequences of facial nerve paralysis, such as the inability to blink, produce facial expressions, difficulty with speech production, and inability to selectively contract facial muscles (synkinesis), which can lead to altered facial expressions (Dalla Toffola et al., 2005; Raslan et al., 2020; Volk et al., 2020). In unilateral facial nerve paralysis, much research effort with transcutaneous FES is dedicated to facial pacing: restoring movements on the paralyzed side of the face based on preserved movements on the intact side of the face. Multiple reports have demonstrated successful application of transcutaneous FES for promoting facial nerve paralysis recovery and symptom management (J. Kim & Choi, 2016; Loyo et al., 2020; Puls et al., 2020; Tuncay et al., 2015, see Burelo-Peregrino et al., 2020; Fargher & Coulson, 2017; Kurz et al., 2022 for reviews). Ongoing research on facial nerve paralysis, its psychological and physical manifestations and clinical FES applications for improving and restoring facial movements can be particularly informative for developing FES systems based on natural verbal and non-verbal communication.

# Complementation of BCI with FES

## BCI-FES for the regulation of stroke and spinal cord injury

Recent research has shown that a combination of a BCI system and a FES system has the potential to create an electronic neural bypass to externally activate muscles affected by paralysis in individuals with stroke and spinal cord injury (Jovanovic et al., 2022; Khan et al., 2020; Marquez-Chin & Popovic, 2020; Silvoni et al., 2011; Sinkjaer et al., 2003; D. M. Taylor, 2008) (**Figure 5**). This can help *improve* motor control by strengthening weakened muscles or *restore* motor control by inducing lost functional movements (Wolpaw et al., 2020). In a combined BCI-FES neuroprosthesis system, neural patterns associated with attempted movements are detected from brain activity, and this information is then used to trigger FES administration in the corresponding muscles. Non-implantable and implantable BCI systems complemented with transcutaneous or implantable FES have been employed to promote rehabilitation in people with stroke and spinal cord injury.

### *Non-implantable BCI-FES*

Most of the research using non-implantable BCI combined with transcutaneous FES systems on stroke patients has focused on the restoration of arm and hand movements. Initial case report data that showed improvements in upper limb motor function for BCI-controlled FES compared to FES-only therapy (Mukaino et al., 2014) was further supported by larger clinical trials (Biasiucci et al., 2018; L. Chen et al., 2021). Although fewer studies have used such BCI-FES systems for post-stroke rehabilitation of lower limb motor function, their findings confirm that BCI-FES systems lead to improvements of motor function (Do et al., 2012; McCrimmon et al., 2014, 2015; Takahashi et al., 2012), and are more effective compared to FES-only therapy (Chung et al., 2015).

A combination of EEG-based BCI and transcutaneous FES was first used to restore hand grasp in an individual with spinal cord injury in the pioneering work by Pfurtscheller et al. (2003). A subsequent study combined an EEG-based BCI with an implanted FES system (G. R. Müller-Putz et al., 2005). More recent studies on BCI-FES for people with spinal cord injury have replicated and extended these findings (Gant et al., 2018; Hernandez-Rojas et al., 2022; Likitlersuang et al., 2018; Muller-Putz et al., 2019), also confirming that combined BCI-FES setups provide significantly greater neurological recovery and functional improvement compared to FES-only therapy (Osuagwu et al., 2016). A recent usability study on combined BCI-FES technology for spinal cord injury rehabilitation showed that users positively evaluated the perceived technology effectiveness, particularly appreciating their apparent active role in the experience (seeing their hand move) and the potential such technology offered for improving interaction with their loved ones (Zulauf-Czaja et al., 2021). This is in agreement with earlier work on user priorities for BCI where user surveys indicated preference of individuals with spinal cord injury for coupling BCI with FES over a BCI that controls a computer, wheelchair or robotic arm (Collinger et al., 2013).

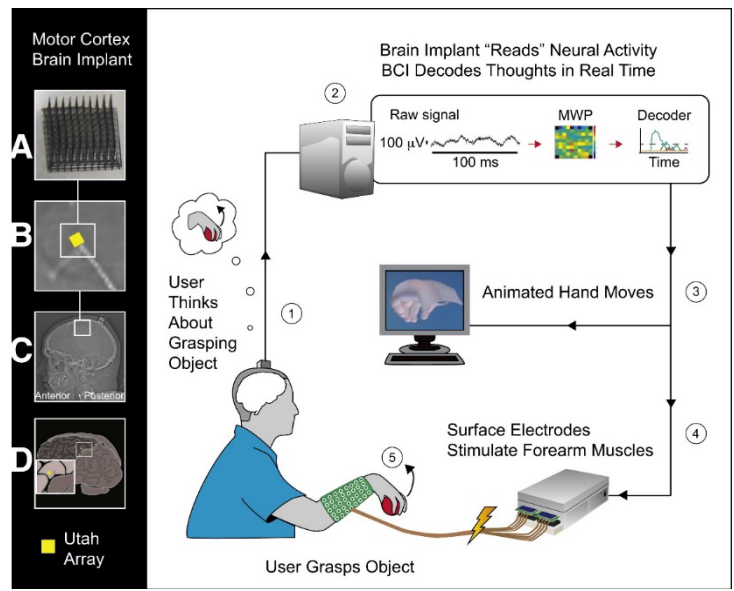

**Figure 5.** Example of a combined BCI-FES system for restoring upper limb movement in an individual with spinal cord injury. Copyright: (Bockbrader, 2019).

*Implantable BCI-FES*

While non-implantable BCI technology provides advantages in terms of availability and safety, implantable BCI technology provides greater neural signal quality in terms of spatial resolution and signal-to-noise ratio (T. Ball et al., 2009). This can potentially allow for decoding of more fine-grained information from the brain about the attempted movement and conceptually could help improve or restore complex continuous control of dynamic functional movements, using FES, as opposed to simple discrete motor actions, such as grasp and release movements.

Recent research has attempted to utilize implantable BCI-FES systems based on ECoG or MEA recordings for the restoration of functional movements in paralyzed individuals. In 2016, a combined application of an implantable MEA-based BCI and transcutaneous FES in a human participant with tetraplegia led to successful restoration of volitional finger, hand, and wrist movements of a paralyzed limb (Bouton et al., 2016). The clinical evaluation results showed that, while using the system, the user's motor impairments significantly improved, which granted the tetraplegic patient with the functional ability to grasp, manipulate, and release objects. These findings were further supported and extended in follow-up research with MEA-based BCI systems combined with FES (Ajiboye et al., 2017; Annetta et al., 2019; Bockbrader, 2019; Colachis et al., 2018; Friedenberg et al., 2017; Sharma et al., 2016). Ajiboye et al. (2017) combined an implantable BCI and a FES device with percutaneous electrodes and an external stimulator to restore functional reaching and grasping movements in an individual with severe tetraplegia, allowing him to repeatedly drink a cup of coffee and feed himself using his own arm and hand, solely on his own volition. Colachis et al. (2018) employed an implantable BCI and transcutaneous FES system that allowed a tetraplegic patient to complete dynamic functional movements using FES-stimulated arm and hand muscles simultaneously with non-paralyzed shoulder and elbow muscles on the same side. Their findings showed that a tetraplegic patient could utilize the system to perform functional tasks ~900 days post implementation, reflecting the strong translational potential of implantable BCI-FES systems for daily life settings. Furthermore, Bockbrader et al. (2019) used an implantable BCI and transcutaneous FES system to restore skillful and coordinated functional grasping movements

(lateral, palmar, and tip-to-tip grips) that provided clinically significant gains in assessment of upper limb functioning in an individual with spinal cord injury (**Figure 5**).

A recent study has presented a combination of an EcoG-based BCI and a transcutaneous FES system for successful restoration of volitional hand grasp in an individual with a spinal cord injury (Cajigas et al., 2021). Finally, a recent fully-implantable EcoG-FES system has been used for restoration of walking in an individual with spinal cord injury (Lorach et al., 2023). The researchers used an implanted spinal stimulation device controlled by an EcoG-based BCI that detected intended movement in bilateral motor cortex neural activity.

**Future BCI-FES research avenues for communication in LIS**

Locked-in individuals experience a loss of motor function, which can cause a total body paralysis. As a result, the affected individuals may be left with little to no means of communication. Augmentative and alternative communication devices and BCI communication systems can offer these individuals alternative ways to communicate. Here, we propose that a technology based on combined BCI-FES systems for the restoration of functional motor activity could represent another research direction that may be worthwhile exploring to provide communication means in the case of LIS.

*Restoring coded communication*

Any intentional natural movement that has been accurately decoded from brain activity and subsequently induced with FES could potentially be used for coded communication and spelling in LIS. Transcutaneous, percutaneous and implanted FES have demonstrated promising results in producing various movements of the hand, arm, leg and face. FES-induced movements can allow for binary "yes-no" and more elaborate multi-class communication modes. This could be coupled with eye trackers, virtual reality and robotics, and potentially new combined technology can be developed.

*Restoring natural communication*

Questionnaires about user needs in assistive communication technology have shown that individuals with LIS have a strong preference for natural personal communication if possible via attempted speech (Branco et al., 2021; Rousseau et al., 2015). In fact, having access only to coded "yes-no" type of communication is associated with lower quality of life compared to richer communication modes (Rousseau et al., 2015). Based on existing work in brain signal decoding and FES-induced body movements, it may also be possible to develop combined BCI-FES technology for more naturalistic communication than coded messages and spelling. For example, a BCI-FES device could allow for pointing and gestures, and fine-grained hand control could make sign language and handwriting possible. Previous BCI work has shown that individual hand gestures, signs of the American Sign Language and handwritten characters can be decoded from neural activity (Branco et al., 2017; Leonard et al., 2020; Li et al., 2017; Pan et al., 2018; Willett et al., 2021), and both transcutaneous and implantable FES could facilitate finger muscle control.

Another potential strategy to communicate can be based on restoration of facial expressions. It has been previously shown that information about facial movements and facial expressions can be decoded from human brain activity for BCI purposes (Bleichner et al., 2015; Salari et al., 2019, 2020). Transcutaneous and percutaneous FES can induce various facial movements, such as eyebrow raises, smiles, and lip puckers, which separately or in combination can lead to perception of a number of facial expressions including anger, happiness, fear, sadness and others (Ekman & Friesen, 2019). This can facilitate rich and natural communication of experienced emotions of the affected individuals with their family and caregivers and have a positive effect on their social interactions overall. Previous studies have indicated that individuals who are unable to produce facial expressions, such as Moebius patients with facial nerve paralysis and persons with LIS, may exhibit deficits in recognizing emotional facial expressions in others (Bate et al., 2013; Calder et al., 2000; De Stefani et al., 2019; Pistoia et al., 2010). Restoring the ability of locked-in persons to produce facial expressions may potentially improve their facial mimicry – an automatic facial reaction that attempts to reproduce perceived emotional facial expression and that is thought to facilitate emotion recognition (Oberman et al., 2007; Ponari et al., 2012; Stel & Van Knippenberg, 2008). FES-induced smiling in FNP patients and individuals with major depressive disorder has been linked to improved quality of life (Demchenko et al., 2023; Lindsay et al., 2014) and might provide similar benefits to persons with LIS.

Although likely unrealistic in the foreseeable future, yet still potentially conceivable, is the idea to decode spontaneous speech from brain activity and use it to trigger orofacial and laryngeal muscles for inducing speech in locked-in individuals. Implantable BCI research has shown impressive results in decoding speech from the brain including individual phonemes, words and even continuous speech segments (Berezutskaya, Saive, et al., 2022; Bocquelet et al., 2016; Chaudhary et al., 2016; Martin et al., 2018; Rabbani et al., 2019 for reviews). Transcutaneous FES research indicates that laryngeal and speech articulators can be activated with external stimulation for producing speech-related motor activity (Kurz et al., 2021; Schultz et al., 2019). Although these FES results are preliminary, they are also promising, as BCI-driven FES-induced speech production has the potential to truly transform the way locked-in individuals could communicate in the future.

Overall, restoring either basic coded communication or more natural communication with combined BCI-FES systems could potentially provide previously inaccessible benefits to users with LIS by offering 1) direct interpersonal communication without having to rely on a computer screen, 2) a natural way to interact with the world, 3) an increased sense of agency and control, 4) somatosensory feedback by design – an important component that has been shown to have positive effects on accuracy and speed of produced movements in BCI users (Flesher et al., 2021).

# Challenges and future directions

There are several challenges and considerations that need to be addressed when testing the feasibility of BCI-FES systems to restore functional movements of body and face in locked-in individuals for the purpose of communication.

## Clinical considerations about LIS

The first challenge relates to the variability of conditions that can potentially lead to LIS. LIS can be caused by various conditions, such as brainstem stroke, trauma, or a progressive neurodegenerative disease (e.g., ALS), and each of them likely differently affects the neuromuscular integrity of limbs and face and therefore the potential success of complementing a communication BCI with FES.

In a situation where paralysis is caused by a disruption of the neural connections between the upper neural pathways and the peripheral and cranial nerves, as is typically the case in brainstem stroke or trauma, it may be feasible to use FES successfully. Notably, individuals with LIS caused by brainstem lesion usually preserve spontaneous involuntary facial expressions while being unable to produce voluntary facial expressions (Hopf et al., 1992; Topper et al., 1995). This dissociation of reflexive and voluntary control may indicate that the facial muscles and peripheral nerves can in principle be excited to produce target expressions but the upper neural pathways involved in voluntary facial movements are impaired. Thus, a BCI-FES combination may help bypass the impaired pathways and restore voluntary facial expressions and potentially other movements that could serve communication in locked-in individuals with a brainstem stroke.

Another important consideration in paralysis that needs to be taken into account is the degree of muscle denervation. Muscle denervation refers to the reduction of nerve inputs into the muscle that causes a decrease in neural input necessary for muscle activation and promotes muscle atrophy. Previous work, however, has shown that transcutaneous FES can be used to induce movement of denervated muscles even several years after the onset of paralysis (Arnold et al., 2021; Kern et al., 2002; Kurz et al., 2022). Activation of denervated muscles with FES may be possible partly due to the process called reinnervation, in which intact neural pathways take over the damaged nerves in controlling muscle activation.

In a neurodegenerative disease, such as ALS, however, the ongoing reinnervation may not be sufficient to preserve functional motor units and as the disease progresses, it will not help compensate for the continually increasing amount of muscle denervation (Hansen & Ballantyne, 1978). Over time, this will inevitably lead to muscle thinning, muscle atrophy, muscle infiltration with fatty tissue and nerve atrophy (Barnes & Simon, 2019; Murray et al., 2010). Limited earlier work has demonstrated successful muscle excitation with electrical stimulation in individuals with ALS (Akaza et al., 2011; Handa et al., 1995), while other work called into question the feasibility and benefits of electrical stimulation in ALS (Amirjani et al., 2012; Herrmann et al., 2022). The stage and progression of the disease may be the determining factor of success in applying FES to restore muscle movement in individuals with ALS with later stages of ALS not being suitable for FES use. More research on muscle denervation in ALS and FES application in denervated muscles is, however, needed to make more specific prognosis about potential outcomes of FES application in LIS caused by a neurodegenerative disease.

In the case where FES cannot successfully induce movements for communication, such as likely during the late stages of ALS, alternative techniques combined with BCI may still apply and benefit from the knowledge gained with existing successful FES applications.

Understanding the neural musculature of body and face, the mapping between cortical neural signals and muscle movements and factors determining successful muscle activation with stimulation could inform novel assistive technology. Such technology could be based on virtual reality, including facial avatars and digital twin development, robotics, orthotics, bionic facial masks and gloves to provide alternative means of communication and interaction with the world for the locked-in individuals.

**Technical considerations of FES**

In the case of translating existing BCI-FES setups to restore motor functioning of upper and lower limb, there may be a number of technical challenges. Optimal choice of stimulation parameters especially in transcutaneous FES needs particular care. Research shows that, in general, paralyzed limbs and face require higher stimulation amplitudes (Mäkelä et al., 2019; Raslan et al., 2020; Volk et al., 2020), which may lead to adverse effects of stimulation. Some studies show that increasing another parameter – pulse duration, may lead to muscle contraction at lower amplitudes (Kurz et al., 2022) and thereby decrease user pain and discomfort (Frigerio et al., 2015; Volk et al., 2020). At the same time, the use of larger pulse durations results in lowered frequency of stimulation. The latter, however, is recommended to be set at 20 – 40 Hz for inducing smooth continuous movements (Doucet et al., 2012; Peckham & Knutson, 2005). Further systematic investigation of optimal parameter configurations may be crucial for acceptability of FES and BCI-FES technologies as a therapeutic tool.

Another challenge in the case of FES applied to facial muscles is the precise electrode placement given the overlapping nature and very small size of facial muscles. Some effort has started on developing recommendations and protocols regarding electrode size and fixation for application of transcutaneous FES to facial muscles (Repitsch & Volk, 2022). Regarding electrode type and placement, transcutaneous facial FES research has relied on existing electromyography guidelines (Fridlund & Cacioppo, 1986). Many studies, however, note the lack of an electromyography site "atlas" available for the facial musculature, and that relevant data is only available for three facial muscles: cheek (zygomaticus major), eyebrow (corrugator supercilia), and forehead (lateral frontalis) muscles. Some researchers are developing facial mapping techniques to identify the facial sites that induces the strongest contraction of the relevant facial muscle (Frigerio et al., 2015), but this practice is not yet widely used in the field.

Another potential challenge is the FES-induced muscle fatigue due to the fact the FES currents may activate muscles in a way and order that are different from naturally induced movements. Specifically, during natural voluntary movement, activation of small fatigue-resistant fibers happens first and then propagates to larger more fatigable muscles, whereas FES may activate larger muscles first or activate muscles non-selectively compromising the natural order of muscle contractions (Feiereisen et al., 1997; Gregory & Bickel, 2005; Vanderthommen & Duchateau, 2007). Animal studies show that stimulation with implanted electrodes may be able to manipulate muscle recruitment order and tackle fatigue (Fang & Mortimer, 1991a, 1991b). With transcutaneous FES, it has been shown that manipulating stimulation parameters can have an effect on fatigue with larger pulse frequencies increasing it (Gorgey et al., 2009; Gregory et al., 2007) and longer pulse durations potentially decreasing it (Jeon & Griffin, 2018). Several studies underline the importance of muscle conditioning and training in reducing fatigue (Guiraud et al., 2014). Despite these efforts, a lot remains unknown about the mechanisms of muscle fatigue and its prevention, and in many cases individual differences between participants continue to determine FES results.

Finally, it is worth considering that using percutaneous and implantable FES electrodes may hold additional benefits for targeted activation of face and body muscles compared to transcutaneous FES. Implantable FES systems may not only be more practical for long-term use, but they may also offer better positioning of the electrodes, do not suffer from issues with electrode-skin impedance on the skin and bypass skin pain receptors. Given that current percutaneous and implantable FES research, especially with facial muscles, shows promising results but remains limited, this could be a potentially noteworthy direction for explorative research on its own and in combination with BCI.

**Practical considerations about combining BCI with FES**

For long-term use, fully-implantable systems may be preferred. This is more practical and esthetically pleasing, which decreases the burden on the user, family, and caregivers. Fully-implantable systems, however, are associated with increased risks of infection, and in the case of a fully-implantable BCI-FES, more components need to be implanted, and therefore a more invasive procedure may be required. In fact, one of the biggest limitations of implantable FES systems is increased risk of infection. These factors need to be carefully considered in designing the systems in order to minimize risks yet maximize benefits for the user.

Several studies have demonstrated successful long-term use of both FES and BCI systems (Agarwal et al., 2003; Colachis et al., 2021; Guiraud et al., 2006, 2014; Milekovic et al., 2018). They show promising results regarding performance and user satisfaction and should be used as guidance in development of combined BCI-FES setups. One important aspect this work can explore is long-term effects of using an assistive technology. In the BCI field, this work focuses on long-term stability of recorded neural control signal and aims to incorporate adaptive data processing methods (Colachis et al., 2021; Rouanne et al., 2022; Shenoy et al., 2006). Both BCI and FES research indicate the importance and challenges associated with training, motivation, and practical implementation of the system (Guiraud et al., 2014; Kleih & Kubler, 2015; Moghimi et al., 2013; Neumann & Kubler, 2003). Work with FES highlights its potential in some cases to counteract muscle denervation and promote muscle growth (Kern et al., 2004; Stein et al., 2010). Studying long-term use of combined BCI-FES technology is necessary to further understand the long-term effects of this technology on the users.

# Conclusions

Recent advances in BCI technology and FES fields have led to the development of combined BCI-FES systems that have demonstrated the feasibility of restoring functional movement of paralyzed limbs in individuals with stroke or spinal cord injury. We anticipate that this research can lead the path to development of novel tools for assistive communication for locked-in individuals as a result of a brainstem stroke, trauma and perhaps neurodegenerative disease, such as ALS. This can be done in two ways: 1) by directly applying existing BCI-FES strategies for upper and lower limb reanimation to locked-in individuals to accomplish basic communication signals, or 2) by exploring a novel direction of reanimating facial movements and expressions with a combined BCI-FES approach. Facial FES research has shown promising results of restoring functional facial movements and expressions in patients with facial nerve paralysis, inducing eyeblinks, eyebrow raises, frowning, smiling. Even limited speech articulation may be conceived for this direction of research. Combined with successful decoding of attempted and intended movements and speech from neural activity of the affected individuals with BCI, we expect a combined BCI-FES approach for providing communication to locked-in individuals to emerge. FES-induced movements and facial expressions driven by neural activity could allow locked-in individuals to express themselves more effectively, which would likely have positive effects for their wellbeing and quality of life.